# THE EVOLUTION OF THE PHASE SPACE DENSITY OF PARTICLE BEAMS IN EXTERNAL FIELDS*

*E.G.Bessonov, Lebedev Physical Institute of the Russian Academy of Sciences, Moscow, Russia*

*Abstract*

In this paper the evolution of the phase space density of particle beams in external fields is presented proceeding from the continuity equation in the six-dimensional (6D) phase space ($\mu$-space). Such a way the Robinson theorem, which includes the Liouville theorem as a special case, was proved in a more simple and consistent alternative way valid for arbitrary external fields, averaged fields of the beam (self-generated electromagnetic fields except intrabeam scattering) and arbitrary frictional forces (linear, nonlinear). It includes particle accelerators as a special case. The limits of the applicability of the Robinson theorem in case of cooling of excited ions having a finite living time are presented.

## INTRODUCTION

In 1958 K.W.Robinson derived at once the sum of damping rates (decrements) of three particle oscillation modes in circular accelerators in the relativistic case [1]. He did an expansion of the power of frictional forces over the particle energy for the private case of radiative reaction force. That is why his final formulae do not include the term with the derivative of the power. Later, A.A.Kolomensky derived the formulae in the general form for the relativistic case and applied it to the ionization cooling [2]. He calculated separately damping rates for three directions in the curvilinear coordinate system and then took their sum.

In order to derive damping increment, K.W.Robinson evaluated the determinant of the transfer matrix of the infinitesimal element of length of a particle orbit. The determinant determines the evolution of a 6D phase space volume of the beam or its density along the trajectory. P.Csonka for this purpose evaluated the infinitesimal 6D phase space volume as well and used some additional conditions to prove the theorem [3]. H.Wiedemann in the textbook [4] presented the proof of the theorem following Robinson's idea but keeping the derivative of the power losses over the energy. Now the theorem in the particle accelerator community is named by Robinson theorem or Robinson's damping criterion.

## EVOLUTION OF PARTICLE BEAM DENSITY IN THE EXTERNAL ELECTROMAGNETIC FIELDS

Let us proceed from the continuity (Liouville's) equation in the 6D phase space coordinate-momentum $(\vec{r}, \vec{p})$:

$$\frac{d\rho}{dt} + \rho \, div \, \vec{v} = 0. \qquad (1)$$

Here components of the 6D velocity $\vec{v} = (\vec{v}_r, \vec{v}_p)$ are $\dot{x}, \dot{y}, \dot{z}, \dot{p}_x, \dot{p}_y, \dot{p}_z$, where $\dot{r}_i = dr_i/dt$, $\dot{p}_i = dp_i/dt$. The equation (1) or equivalent equation $\partial \rho / \partial t + div \, (\rho \vec{v}) = 0$ expresses the number of particles conservation law. In our case the form of the equation (1) is preferable as it presents the total derivation of the density in the coordinate system moving with the beam. It can be presented in the integral form $\rho = \rho_0 \exp[-\int div(\vec{v}) dt]$, where $\rho_0$ is the initial phase space density.

The divergence $div \, \vec{v} = div_r \vec{v}_r + div_p \vec{v}_p$, $div_r \vec{v}_r = 0$ as the velocity $\vec{v}_r = c\vec{p}/\sqrt{p^2 + m^2 c^2}$ does not depend on spatial coordinates ($r_i$, $p_i$ are independent variables). The value $\vec{v}_p = \dot{\vec{p}} = \dot{\vec{p}}_H + \dot{\vec{p}}_{Fr} = \vec{F}_H + \vec{F}_{Fr}$ is the force acting upon the particle. The conservative force $\vec{F}_H = e\vec{E}(\vec{r},t) + (e/c)[\vec{v} \, \vec{H}(\vec{r},t)]$ is determined by external fields and the fields of the particle beam ($div_p \vec{F}_H = 0$), while $\vec{F}_{Fr}$ is the frictional force. That is why $div_p \vec{v}_p = div_p \vec{F}_{Fr}$, and the equation (1) can be presented in the integral form $\rho = \rho_0 \exp[-\int div_p \vec{F}_{Fr} dt]$.

The frictional force can be written in the form $\vec{F}_{Fr} = -\chi_{Fr}(\vec{r}, p, t) \cdot \vec{n}$, where $p = |\vec{p}|$, $\vec{n} = \vec{p}/p$, $\chi_{Fr}(\vec{r}, p, t)$ is the frictional coefficient. In this case $div_p \vec{F}_{Fr} = -\chi_{Fr} div_p \vec{n} - \vec{n} \cdot grad_p \chi_{Fr} = -2\chi_{Fr}/p - \partial \chi_{Fr}/\partial p$. We took into account, that $div_p \vec{n} = 2/p$ and $\vec{n} \cdot grad_p \chi_{Fr}(\vec{r}, p, t) = (\partial \chi_{Fr}/\partial p) = v_r (\partial \chi_{Fr}/\partial \varepsilon)$, where $\varepsilon = \sqrt{p^2 c^2 + m^2 c^4}$ is the energy of the particle.

The frictional power $P_{Fr} = \vec{F}_{Fr} \cdot \vec{v}_r = \chi_{Fr}(\vec{r}, p, t) \cdot \vec{n} \cdot \vec{v}_r = c\beta \cdot \chi_{Fr}(\vec{r}, p, t)$, where $\beta = v_r/c$. It follows that $\chi_{Fr} = P_{Fr}(\vec{r}, p, t)/c\beta$, and the equation (1) become

$$\rho = \rho_0 \exp[-\int \alpha_{6D}(\vec{r}, p, t) \, dt], \qquad (2)$$

where $\alpha_{6D}(\vec{r}, p, t) = -div_p \vec{F}_{Fr} = 2\chi_{Fr}/p + \partial \chi_{Fr}/\partial p$ or:

$$\alpha_{6D}(\vec{r}, \varepsilon, t) = (1 + \frac{1}{\beta^2})\frac{P_{Fr}(\vec{r}, \varepsilon, t)}{\varepsilon} + \frac{\partial P_{Fr}(\vec{r}, \varepsilon, t)}{\partial \varepsilon}.$$

The integral (2) along a trajectory of a particle is the solution of the equation (1). According to (2), the 6D rate of the beam density change is determined by the frictional power and its derivative with respect to the particle energy. The integral (2) is valid for the arbitrary systems (linear, nonlinear, coupled). We did not use a curvilinear coordinate system, the Jacobee's formula for the system of linear differential equations, matrices; any additional conditions (see [3]). In our case the expression (2) is valid for the nonrelativistic case as well. In general case ($\alpha_{6D} \neq const$) the solution is not exponential function. The equation (2) is valid for infinitesimal parts of the beam. The damping of different parts of the beam in some methods of cooling should be distinguished.

If $P_{Fr}(p) = 0$, then $\alpha_{6D}(\vec{r}, p, t) = 0$, $d\rho/dt = 0$ and, according to (2), we come to the Liouville's theorem [4]–[7]. It states that for conservative systems the particle density $\rho$ in the 6D phase space, the number of particles in the phase space volume occupied by the beam and hence the volume stay constant. In this case the volume is named the 6D normalized emittance. The volume divided by the factor $(\beta\gamma)^3$ is named the unnormalized emittance accordingly. The normalized emittance is invariant for conservative systems. 2D and 4D phase space volumes can be exchanged by conservative external fields. It follows from private examples (see e.g. [1]).

Robinson and Liouville theorems are valid for identical particles (electrons, protons, muons and so on). The Robinson theorem is valid if frictional forces exist only at the moment of their interaction with media or external fields and there is no time delay between the interaction time and the frictional force. Excited ions have higher rest mass then unexcited ones and have finite lifetime. It means that in general case the above theorems are not valid for ion cooling (excited ions are not identical to unexcited ones and have finite lifetime). The theorems works well if the lifetime of excited ions is less then some characteristic time for the processes of the beam evolution determined by concrete conditions. E.g., in the case of particle accelerators the delay time between the moment of the ion excitation by a laser beam and following photon reemission must be less then the period of betatron oscillations. Otherwise, the additional cooling or heating of the ion beam is possible.

It is supposed above that the particle beam is a continuous media (there is no free space between particles). On practice it means that the distance between particles is much less then dimensions of any instrument, which is used to move a particle from a peripheral region of the 6D volume of the beam to the central one and that the instrument do not disturb another particles of the beam (otherwise, the stochastic cooling is possible).

## APPLICATION OF ROBINSON THEOREM TO PARTICLE ACCELERATORS

The equation (2) can be applied to the beam transport lines, linear and circular accelerators, storage rings, betatrons, recirculators and so on. If the motion of particles of a beam in the limits of the occupied 6D volume is described by linear differential equations, then all parts of the beam density and the total phase space volume occupied by the beam change identically.

If the energy losses are compensated by induction or RF fields in cavities of the circular machines, the average energy of particles is kept constant, power loss of particles of the beam in the limits of its phase space volume is a linear function of the energy, then, according to (2), the rate of the beam density change in these machines is determined by the 6D-damping increment

$$\overline{\alpha_{6D}} = -\overline{div\ \vec{F}_{Fr}} = (1+\frac{1}{\beta^2})\frac{\overline{P_{Fr}(\varepsilon)}}{\varepsilon}\Big|_{\varepsilon=\varepsilon_s} + \frac{\partial \overline{P_{Fr}(\varepsilon)}}{\partial \varepsilon}\Big|_{\varepsilon=\varepsilon_s}, (3)$$

where $\overline{P_{Fr}}(\varepsilon)$ is the average rate of the particle energy loss due to friction, $\varepsilon_s$ is the energy of the reference particle. Reference particle can be synchronous particle if the radiofrequency accelerating field is switched on or some central particle if the induction accelerating field is used. If the increment $\alpha_{6D}$ is positive then in this case they say that 6D cooling occurs. The expression (3) can be negative, if the second term $\partial \overline{P}_{Fr}(p)/\partial \varepsilon$ is negative and larger by the value then the first one (beam heating).

Note that 6D cooling is a necessary but not sufficient condition for production of the high density beam in 3D space. Untidamping of synchrotron [8] or betatron [9] oscillations can occur under 6D cooling conditions. The emittance exchange can be used to remove the case of untidamping [1], [4], [9].

The following kinds of the charged particle energy looses can be used for cooling: electromagnetic radiation losses in the external electromagnetic fields (synchrotron radiation, undulator radiation, backward Compton and backward Rayleigh scattering) and in matter (ionization and excitation losses, bremstrahlung radiation).

If the energy of particles is maintained at constant level, $\overline{P_{Fr}}(p) \neq 0$, and $\partial \overline{P_{Fr}}(p)/\partial \varepsilon \sim \overline{P_{Fr}}(p)/\varepsilon$ then, according to (3), the density of the particle beam will increase by $e = 2.7$ times after particles of the beam will lose the energy $\Delta\varepsilon \sim \varepsilon$ (e.g., synchrotron radiation damping). If $\partial \overline{P_{Fr}}(p)/\partial \varepsilon \sim \overline{P_{Fr}}(p)/\Delta\varepsilon_b$ ($\Delta\varepsilon_b \ll \varepsilon$) then the same increase of the density will be after particles of the beam lose the energy $\Delta\varepsilon = \Delta\varepsilon_b$, where $\Delta\varepsilon_b$ is the initial energy spread of the beam. This case is possible for the radiative cooling of ion beams by the broadband laser beam [10]. In this case the power loss must increase with the energy from zero to maximal value in the limits of the energy spread $\Delta\varepsilon_b$ (fast, enhanced, stimulated cooling) [11]-[12].

To separate the longitudinal component $p_s$ of the momentum from the transverse components $p_x$, $p_z$ we must find the closed trajectory of a reference particle, direct the longitudinal unit vector along the trajectory and direct two other transverse unit vectors in the directions transverse to each other and to the longitudinal one. In such a way we will pass on to a curvilinear coordinate system for particle accelerators. In this system the 6D increment is the sum of two transverse (radial, vertical) and longitudinal increments: $\alpha_{6D} = 2\alpha_x + 2\alpha_z + 2\alpha_s$. The increments for longitudinal and uncoupled vertical oscillations are found without a problem by direct calculations and the radial one is determined from the equation (3). In the relativistic case:

$$\alpha_x = \frac{1}{2}\left[\frac{\overline{P}_s}{\varepsilon_s} + \frac{\partial \overline{P}}{\partial \varepsilon}|_s - \frac{d\overline{P}}{d\varepsilon}|_s\right], \quad \alpha_z = \frac{1}{2}\frac{\overline{P}_s}{\varepsilon_s}, \quad \alpha_s = \frac{1}{2}\frac{d\overline{P}}{d\varepsilon}|_s. \quad (4)$$

The total momentum and the energy of a particle are coupled with its components $p_x$, $p_z$, $p_s$ by the relations $p^2 = p_x^2 + p_z^2 + p_s^2$, $\varepsilon = p^2 + mc^2$. That is why we could pass from $\mu$-space to the new space ($x$, $p_x$, $z$, $p_z$, $s$, $\varepsilon$), which is usually used in the theory of accelerators. We can name it the $\varepsilon$-space. Particles in the accelerators and storage rings are subjected to fast transverse vertical and radial betatron oscillations. Moreover they participate in longitudinal oscillations of two types: fast oscillations of momentum with the betatron frequencies in accordance with the equation $p_s^2 = p^2 - p_x^2 - p_z^2$ and, if the accelerating radiofrequency field is switched on, slow synchrotron oscillations of the momentum $p(t)$ and energy $\varepsilon(t)$.

Note that frictional cooling goes through the energy losses of particles, particle beam can stay monoenergetic, its energy spread and 6D volume are equal to zero (particle beam is on the 6D hypersphere), but the spread of the longitudinal oscillations of particles will not be zero if the transverse emittances are nonzero.

## CONCLUSION

The evolution of the phase space density of particle beams in external fields is presented proceeding from the continuity equation in the six-dimensional phase space. The Robinson theorem, was proved in an alternative way valid for arbitrary external fields, averaged fields of the beam (self-generated electromagnetic fields except intrabeam scattering) and arbitrary frictional forces. It includes particle accelerators as a special case. The limits of the applicability of the Robinson theorem in case of cooling of excited ions having a finite living time are presented.

The author thanks R.M.Feshchenko for usefull discussions.